\documentclass[aps,pre,reprint,amsmath,amssymb]{revtex4-1}
\usepackage[pdftex]{graphicx}
\usepackage{subfigure}
\usepackage{dcolumn}
\usepackage{epstopdf}
\usepackage{color}
\usepackage{upgreek}
\usepackage[hidelinks]{hyperref}

\usepackage{bm}

\makeatletter
\def\btt#1{\texttt{\@backslashchar#1}}%
\DeclareRobustCommand\bblash{\btt{\@backslashchar}}%
\makeatother

\begin{document}

\title{ Non-monotonous shear rate dependence of dielectric relaxation frequency of a nematic liquid crystal revealed by rheo-dielectric spectroscopy  }

\date{\today}
\author{K. Anaswara Das, M. Praveen Kumar and Surajit Dhara}
\email{surajit@uohyd.ac.in} 
\affiliation{School of Physics, University of Hyderabad, Hyderabad-500046, India\\ }

\begin{abstract}
 Dielectric relaxation of materials provides important information on the polarisation dynamics at different time scales. We study the dielectric relaxation of a nematic liquid crystal under steady rotational shear and simultaneously measure the viscosity. The dielectric anisotropy of the nematic is positive and the applied electric field is parallel to the velocity gradient direction with a magnitude larger than the Freedericksz threshold field. The complex dielectric constant and effective viscosity decrease rapidly with increasing shear rate. The dielectric relaxation frequency exhibits a non-monotonous shear rate dependence, first decreasing but beyond a critical shear rate increasing. 
 Our experiments suggest the emergence of collective dipolar relaxation under the influence of the competing effects of hydrodynamic and dielectric torques.

 \end{abstract}
\preprint{HEP/123-qed}
\maketitle

\section{Introduction}
The structures and properties of complex fluids such as polymeric systems, colloidal suspensions and liquid crystals have been studied under simultaneous shear flow and electric fields~\cite{lar,mp}. In some fluids, the viscosity is enhanced significantly due to the application of an electric field. This effect is known as the electrorheological effect and is very important for applications in electromechanical devices~\cite{www,gamo,fuka,djk}. 
In nematic liquid crystals (LCs), composed of axially polar molecules, under the application of sufficient electric field the effective viscosity is enhanced and this effect is due to the change in the orientation of the director (average aligning direction of the molecules) with respect to the shear flow direction~\cite{TC,SS,K-L, AD,negita,hw,patrico,MCT}. 
In uniaxial nematic LCs, there are two principal dielectric constants, namely $\epsilon_{||}$ and $\epsilon_{\perp}$, where the subscripts refer components in relation to the director~\cite{pg}. The dielectric anisotropy ($\Delta\epsilon =\epsilon_{||}-\epsilon_{\perp}$) of the nematic LCs composed of polar molecules could be zero, positive or negative, depending on the orientation of the permanent dipole moments with respect to the long molecular axis~\cite{deju}. 

Both the dielectric constants are frequency dependent and the corresponding relaxation frequencies $f_{||}$ and $f_{\perp}$ are related to the time scales of rotation of the dipoles along the short and long molecular axes, respectively~\cite{deju}. For axially polar molecules, the relaxation frequency $f_{||}$ is much smaller than $f_{\perp}$ and it is the opposite in the case of transversely polar molecules~\cite{deju}. Usually, these relaxation frequencies are measured in cells composed of parallel electrodes in which the director is aligned either perpendicular (homeotropic) or parallel (homogeneous) to the electrodes. In cells the nematic LCs are in the quiescent state, hence such measurements provide equilibrium relaxation times which are important for device applications. 
 	
 A nematic LC is far away from the equilibrium when subjected to a steady shear flow and depending on the system it may exhibit three dynamic modes, namely flow-aligned, wagging and tumbling~\cite{fmles1,fmles2,fmles3,andre,grsh,vvb}. In flow aligned nematics, the director forms a steady state with a tilt angle $\theta_L$ (Lesli angle) with respect to the flow direction whereas, in the case of wagging, the director oscillates and for tumbling motions, the director makes full rotations in the shear plane~\cite{aarch,djter,wrbu,scho,ygta}. 
 In a flow-aligned nematic with positive dielectric anisotropy ($\Delta\epsilon > 0$) the application of sufficient electric field perpendicular to the flow direction induces an apparent change in viscosity~\cite{pg}. The change in viscosity due to the change in the director orientation can be measured easily from the rheo-dielectric studies at a frequency much below the dielectric relaxation frequency~\cite{negita,patrico,MCT,JA1,JA2}.  However, so far it is not well understood if the shear flow affects the dielectric relaxation frequencies of the nematic LCs~\cite{HW1,HW2}.
  Since the shear rate is much slower than the equilibrium rotation of the dipoles ($\dot{\gamma} \ll \tau^{-1}$, relaxation time), one would expect that the shear flow should not affect the dielectric relaxation frequency. The rheo-dielectric studies at low electric fields on some cyanobiphenyl nematic LCs show no measurable effect of shear flow on the dielectric relaxation frequency~\cite{HW1}.  \\
 \indent In this paper, we apply a sufficiently high electric field that orients the director of the nematic along the velocity gradient direction (perpendicular to the shear flow direction) and study the frequency dispersion of the effective dielectric constant of a nematic liquid crystal at various shear rates.
 We measure the real and imaginary parts of the dielectric constants and analyse their frequency dependence at varying voltage and shear rates. Our study shows evidence of a collective dielectric relaxation mode in the intermediate shear rate range. We discuss the possible origin of the collective dipolar relaxation.

\section{Experiment}
\begin{figure}[!ht]
	\center\includegraphics[scale=0.32]{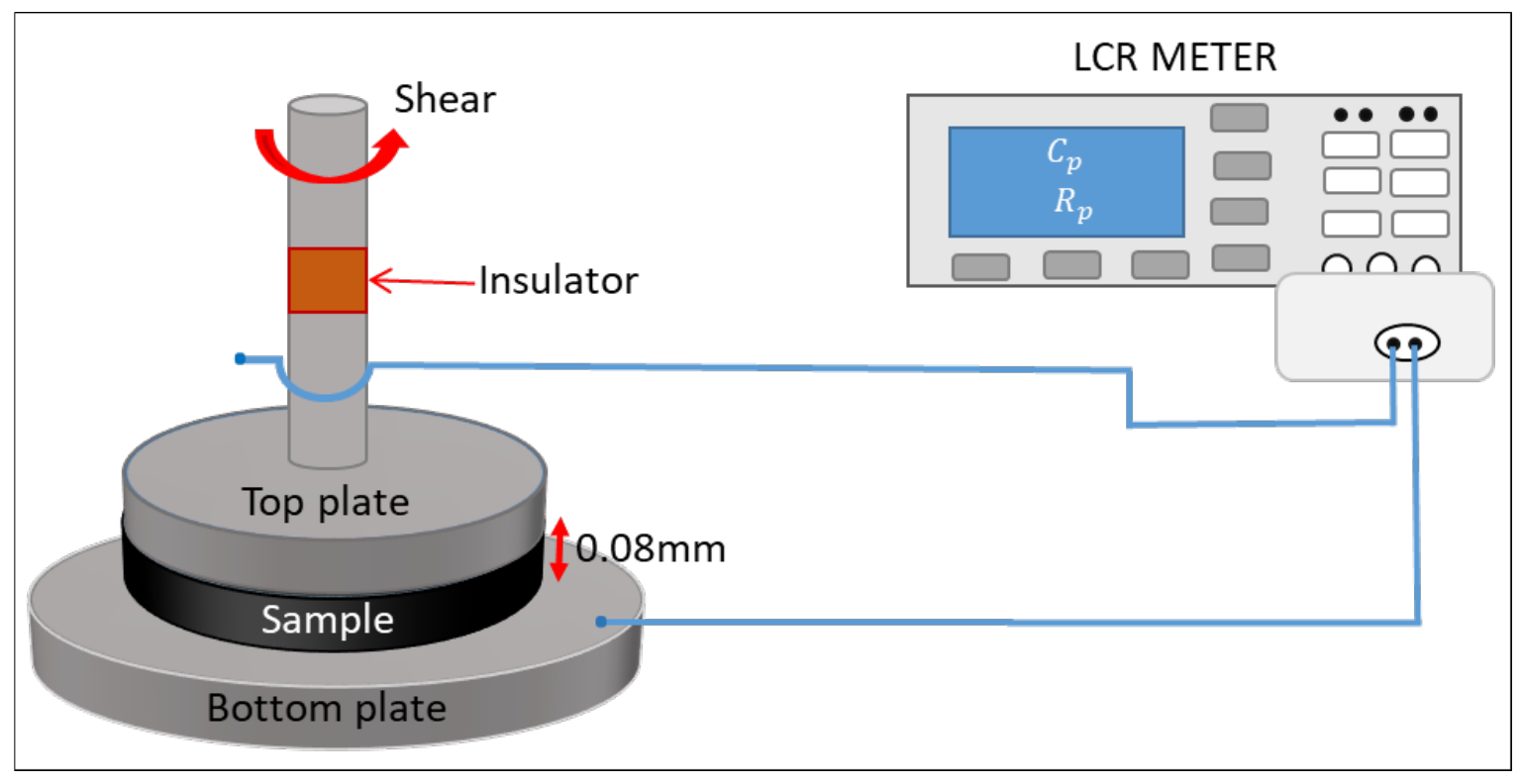}
	\caption{Experimental setup for rheo-dielectric measurements.
	 \label{fig:figure1}}
\end{figure}

We worked on a nematic liquid crystal mixture, known as E7. It is a mixture of several cyanobiphenyl, cyanoterphenol and triphenyl compounds at some specific compositions~\cite{synthon}. It was obtained from Grand Winton Inc, and used without further purification. At  5$^\circ$C the parallel and perpendicular components of the dielectric constant are given by $\epsilon_{||}=19.6$ and $\epsilon_{\perp}=4.5$, respectively and the dielectric anisotropy is positive ($\Delta\epsilon=\epsilon_{||}-\epsilon_{\perp}$ = 15.1)~ (see Fig.S1, SM)\cite{x2}. The nematic to isotropic phase transition temperature is 61 $^{\circ}$C and it exhibits a broad nematic temperature range (61$^\circ$C to $-10^\circ$C). The rheo-dielectric measurements were made by a strain-controlled rheometer (Anton Paar MCR 501) using parallel-plate geometry with a plate diameter of 50 mm. The plate gap was fixed at 80$~\mu$m and no alignment layer was used in this experiment. The bottom plate was fixed while the top plate was rotated at different shear rates. The bottom plate was attached with a Peltier temperature controller for controlling the temperature with an accuracy of $0.1^\circ$C. A hood was used to cover the measuring plates for temperature uniformity. An LCR meter (Agilent E4980A) was connected to the plates for rheo-dielectric measurements. The applied electric field was greater than the Freedericksz threshold field ($E\gg E_{th}$) and the frequency was increased from 100 Hz to 1.0 MHz. The experiments were performed at different shear rates and the frequency-dependent real and imaginary parts of the dielectric constant were measured by the LCR meter.
All the measurements were made at 5$^{\circ}$C.  Before the measurement, the sample was pre-sheared for 5 minutes at a fixed shear rate ($\dot{\gamma}$=100 s\textsuperscript{-1}). A schematic diagram of the experimental setup is shown in Fig.\ref{fig:figure1}. 

\section{Results and discussion}

\begin{figure}[!ht]
	\center\includegraphics[scale=0.25]{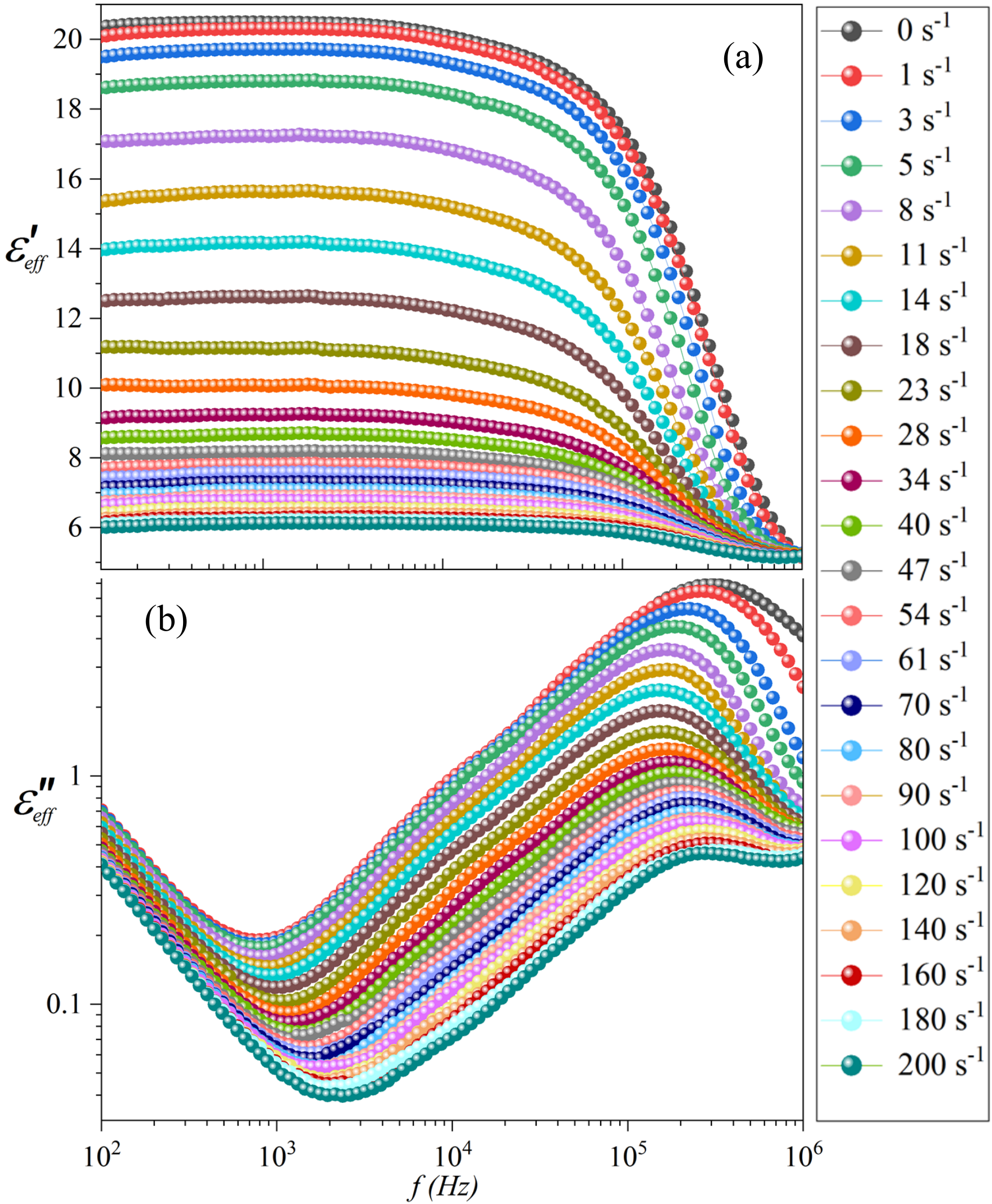}
	\caption{Plots of (a) real 
	($\epsilon_{eff}^{\prime}$) and (b) imaginary ($\epsilon_{eff}^{\prime\prime}$) components of the effective dielectric constant with frequency $f$, at different shear rates and temperature T=5$^\circ$C. The applied electric field $E=2.5\times10^5 V/m$, which is 10 times higher than the Freedericksz threshold field, $E_{th}=2.5\times10^4~V/m$. 
	 \label{fig:figure2}}
\end{figure}

We measured the frequency dispersion of the effective dielectric constants at a fixed electric field and different shear rates (Fig.\ref{fig:figure2}). At room temperature the complete relaxation mode is not observed due to the limiting frequency range of the LCR meter (100 Hz to 2 MHz) hence, all the experiments were performed at a fixed temperature of 5 $^\circ$C. 
Figure \ref{fig:figure2}(a,b) shows the dielectric dispersion of the real ($\epsilon_{eff}^{'}$) and imaginary ($\epsilon_{eff}^{''}$) parts of the effective dielectric constant at different shear rates.  The applied electric field was $E_{th}=2.5\times10^5~V/m$, which is 10 times larger than the Freedericksz threshold field ($E_{th}=2.5\times10^4~V/m$).
   In the low-frequency region, ($f < 10^{4}$ Hz), at zero shear rate (i.e., quiescent nematic), $\epsilon_{eff}^{'}$ is independent of frequency and decreases with increasing shear rate. On the other hand, $\epsilon_{eff}^{''}$ decreases with the increasing frequency in the low-frequency region ($f < 10^{3}$ Hz). It varies as $\epsilon^{''}_{eff}\propto f^{-1}$, which suggests 
  the influence of the conductivity (direct current) on the imaginary component of the dielectric constant.

\begin{figure}[!ht]
\center\includegraphics[scale=0.18]{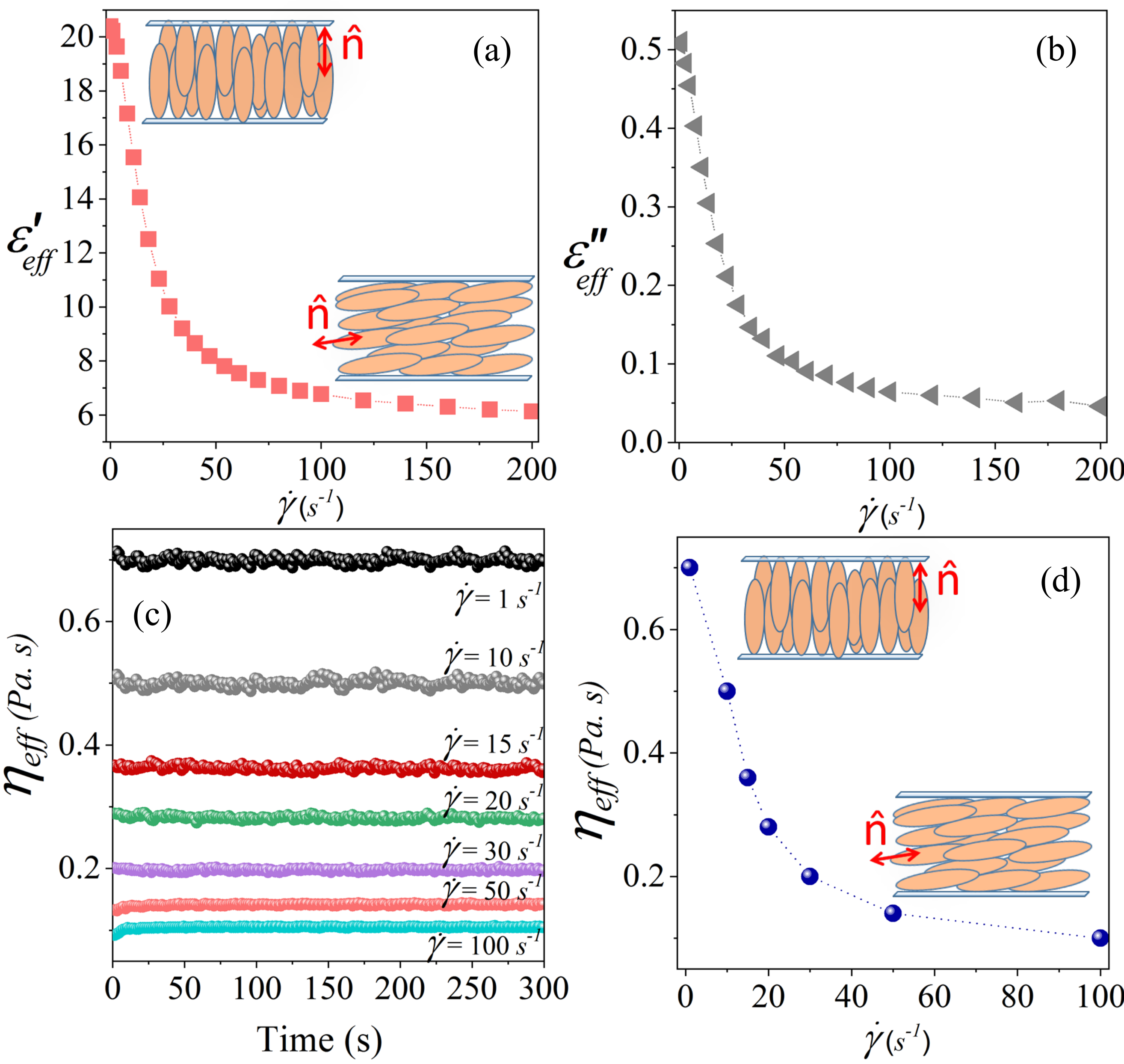}
\center\caption{ Variation of (a) $\epsilon_{eff}^{\prime}$ and (b) $\epsilon_{eff}^{\prime\prime}$ with shear rate at frequency $f=4$ kHz (obtained from Fig.\ref{fig:figure1}). (c) Time-dependent effective viscosity $\eta_{eff}$ at $E=2.5\times10^5 ~V/m$ and at different shear rates. (d) Variation of $\eta_{eff}$ at different shear rates obtained from the figure (c). Shear-dependent director orientations are shown in the insets.
\label{fig:figure3}}
\end{figure}
 
 Figure \ref{fig:figure3}(a) and (b) shows the variation of $\epsilon^{'}_{eff}$ and $\epsilon^{''}_{eff}$ obtained from Fig.\ref{fig:figure1} with the shear rate at a fixed frequency $f=4$ kHz. Both decrease exponentially with increasing shear rate. For example, at $\dot{\gamma}=0$ s\textsuperscript{-1}, $\epsilon^{'}_{eff}=20.3$. This value is nearly equal to the parallel component of the dielectric constant measured in a homeotropic cell \textit{i.e.,} $\epsilon_{||}\simeq19.6$~ (see Fig.S1, SM)\cite{x2}. It suggests that the director is almost perpendicular to the confining plates and parallel to the electric field direction. With increasing shear rate, $\epsilon^{'}_{eff}$ decreases rapidly and eventually becomes constant \textit{i.e,} $\epsilon^{'}_{eff}\simeq6$, when the shear rate is increased to $\dot{\gamma}=200$ s\textsuperscript{-1}. This dielectric constant at this shear rate is slightly larger than the value measured in a homogeneous (planar) cell \textit{i.e.,} $\epsilon_{\perp}\simeq4.5$ (see Fig.S1, SM)\cite{x2}. Hence, these results demonstrate that initially ($\dot{\gamma}=0$ s\textsuperscript{-1}) the director is perpendicular to the plates \textit{i.e.,} parallel to the velocity gradient direction and it gradually tilts in the shear plane towards the shear flow direction. Further, the effective viscosity $\eta_{eff}$ of the liquid crystal was measured simultaneously with the dielectric dispersion measurements. Figure \ref{fig:figure3}(c) shows that the viscosity is independent of time but it decreases with increasing shear rate. Figure \ref{fig:figure3}(d) shows the variation of the viscosity at different shear rates. At $\dot{\gamma}=1$ s\textsuperscript{-1}, $\eta_{eff}=0.7$ Pa s and it reduces to about $0.11$ Pa s when $\dot{\gamma}=100$ s\textsuperscript{-1}. At $\dot{\gamma}=1$ s\textsuperscript{-1}, the effective viscosity can be considered as the Miesowicz viscosity for the director orientation being parallel to the velocity gradient direction \textit{i.e.,} $\eta_{eff}\simeq\eta_{1}$ and at the highest shear rate ($\dot{\gamma}=100$ s\textsuperscript{-1}),  $\eta_{eff}\simeq\eta_{3}$, where $\eta_3$ is the Miesowicz viscosity with the director being nearly parallel to the velocity direction.

One important observation is the change of dielectric relaxation frequency with shear rate (Fig. \ref{fig:figure2}(b)). It is apparent that not only does the peak height of the $\epsilon^{''}_{eff}$ decrease, but the frequency at which the peak occurs, so-called the dielectric relaxation frequency also changes with shear rate. To obtain the actual dielectric relaxation frequency ($f_r$), we fitted the dielectric data to the Havriliak-Negami relaxation function~\cite{hane},
	
\begin{equation}
\epsilon^{*}(f)= \epsilon_{\infty}+\frac{\Delta\epsilon}{[1+(i 2\pi f\tau)^{\alpha}]^{\beta}}-i\frac{\sigma_{0}}{\epsilon_{0} 2\pi f}
	\end{equation}
where $\Delta\epsilon$ is the dielectric strength, $\epsilon_{\infty}$ is the dielectric permittivity at the high-frequency limit, $\sigma_{0}$ is the conductivity and $\tau$ is the relaxation time. The corresponding relaxation frequency is given by $f_r = 1/2 \pi \tau$ (see Fig.S2, SM~\cite{x2}). The exponents $\alpha$ and $\beta$ describe the asymmetry and broadness of the corresponding spectra and are found to vary in the range of 0.8 to 1 (see Table S1, SM~\cite{x2}).

\begin{figure}[htbp]
\center\includegraphics[scale=0.25]{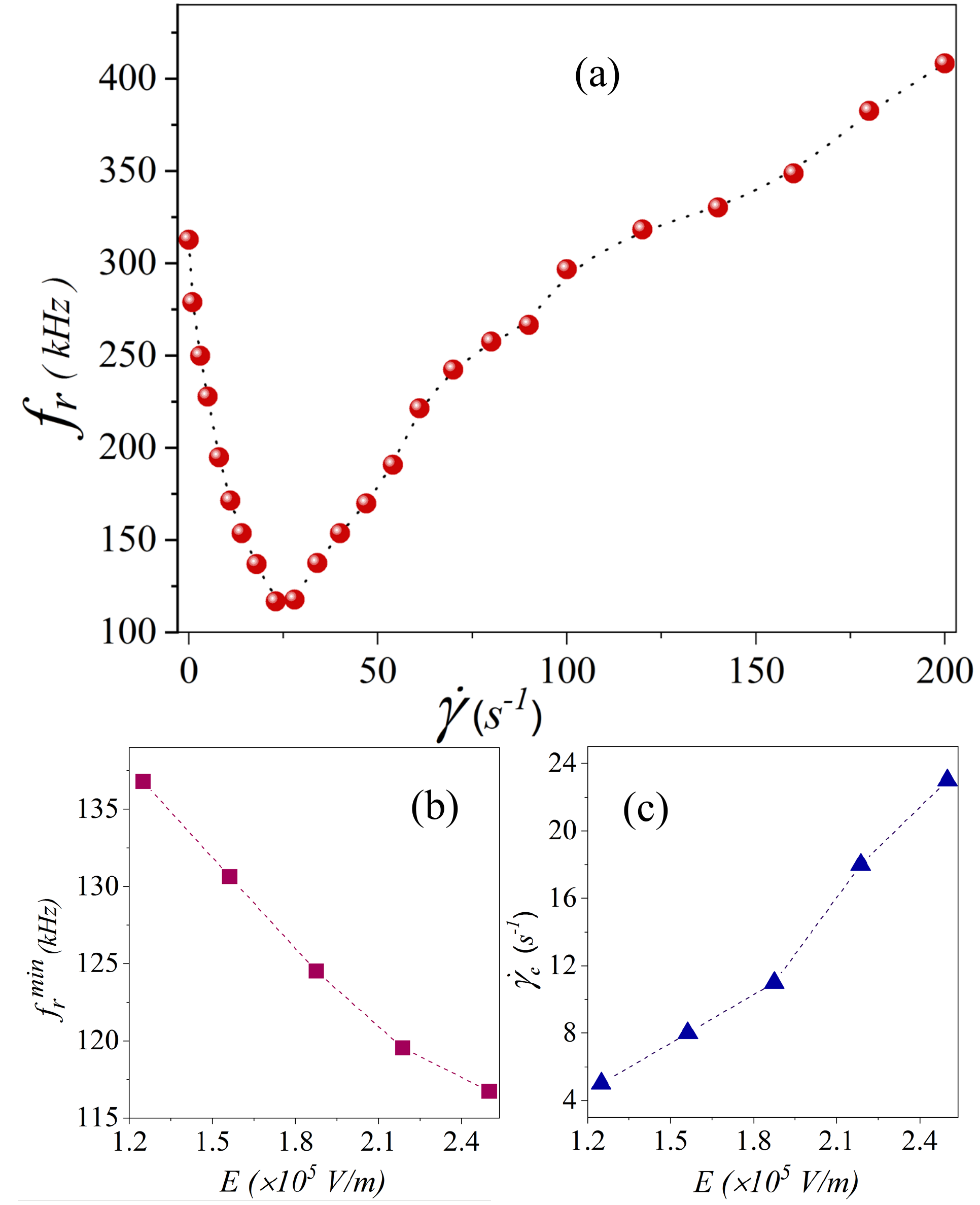}
\center\caption{ (a) Variation of the effective relaxation frequency $f_{r}$ with shear rate at $E=2.5\times10^5 ~V/ m$. Dependence of (b) minimum relaxation frequency $f^{min}_r$ and (c) the critical shear rate $\dot{\gamma_c}$.  The standard deviation of the relaxation frequency at zero shear rate is about 2 kHz.
\label{fig:figure4}}
\end{figure}

 Figure \ref{fig:figure4}(a) shows the variation of the relaxation frequency $f_r$ with the shear rate $\dot{\gamma}$ at an applied field of $E=2.5\times10^5 ~V/m$. The relaxation frequency decreases from 320 kHz to 116.7 kHz when $\dot{\gamma}$ is increased from 0 to 23 s\textsuperscript{-1}. It shows a pronounced minimum at $\dot{\gamma}=23$ s\textsuperscript{-1}, above this, $f_r$ increases to about 400 kHz, when $\dot{\gamma}$ is further increased to 200 s\textsuperscript{-1}. 
  At $\dot{\gamma}=0$ s\textsuperscript{-1}, the director is parallel to the field direction, hence the effective relaxation frequency $f_r$ corresponds to the rotation of the longitudinal (axial) components of the dipole moments about the short axis \textit{i.e.,} $f_r \sim f^{||}_{r}\sim 320$ kHz, where $f^{||}_{r}$ is the relaxation frequency measured in homeotropic cell (quiescent nematic)(see Fig.S1, SM)~\cite{x2}.
The relaxation frequency of the transverse dipole moments about the long axis in the homogeneous cell (quiescent nematic) is $f^{\perp}_{r}> 2$ MHz (see Fig.S1, SM)~\cite{x2} which is much larger than $f^{||}_{r}$.  Hence, one would have expected the relaxation frequency $f_r$ to increase continuously from 320 kHz to 400 kHz with increasing shear rate. Thus, the minimum relaxation frequency ($f^{min}_r$) at a particular shear rate is rather unexpected and can not be explained simply by considering the tilting of the director. In this context, two possible director dynamics should be mentioned. Any effect of electrohydrodynamic instability on the dielectric relaxation is ruled out as both the dielectric and conductivity anisotropies ($\Delta\epsilon$ and $\Delta\sigma$) are positive~\cite{blinov}. Further, E7 is a flow-aligning nematic hence the tumbling and wagging of the director is absent~\cite{x1}. Hence, the reduction of relaxation frequency, in other words, the slowing down of the relaxation time ($\tau$) could hint at a possible collective dipolar relaxation. One can expect that the collective response should also contribute to enhancing the effective dielectric constant. However, such an effect can not be observed because of the counter effect of tilting of the director in the shear plane that tends to reduce the effective dielectric constant. 
  
We performed experiments at a few different electric fields ($E\gg E_{th}$) at the same temperature~(see Fig.S3 and S4, SM \cite{x2}) and observed that the minimum relaxation frequency $f^{min}_r$ decreases almost linearly with the increasing field (Fig. \ref{fig:figure4}(b)).  The critical shear rate $\gamma_c$ at which the relaxation frequency is minimum is also found to increase with the field (Fig. \ref{fig:figure4}(c)). We measured the dielectric relaxation frequency at a few electric fields in a homogeneous cell of thickness $11.3~\mu m$ (quiescent nematic). It is observed that the relaxation frequency decreases almost linearly with increasing electric field as expected (see Fig.S5, SM~\cite{x1}). Hence, the decrease of $f^{min}_r$ with increasing field under shear flow can be attributed to the increasing tilt angle of the molecules with respect to the field direction. Figure \ref{fig:figure4}(c) demonstrates the competing effects, i.e., the high shear rate is required at a larger electric field for the collective response to be observed (see later discussion).

\begin{figure}[htbp]
\center\includegraphics[scale=0.16]{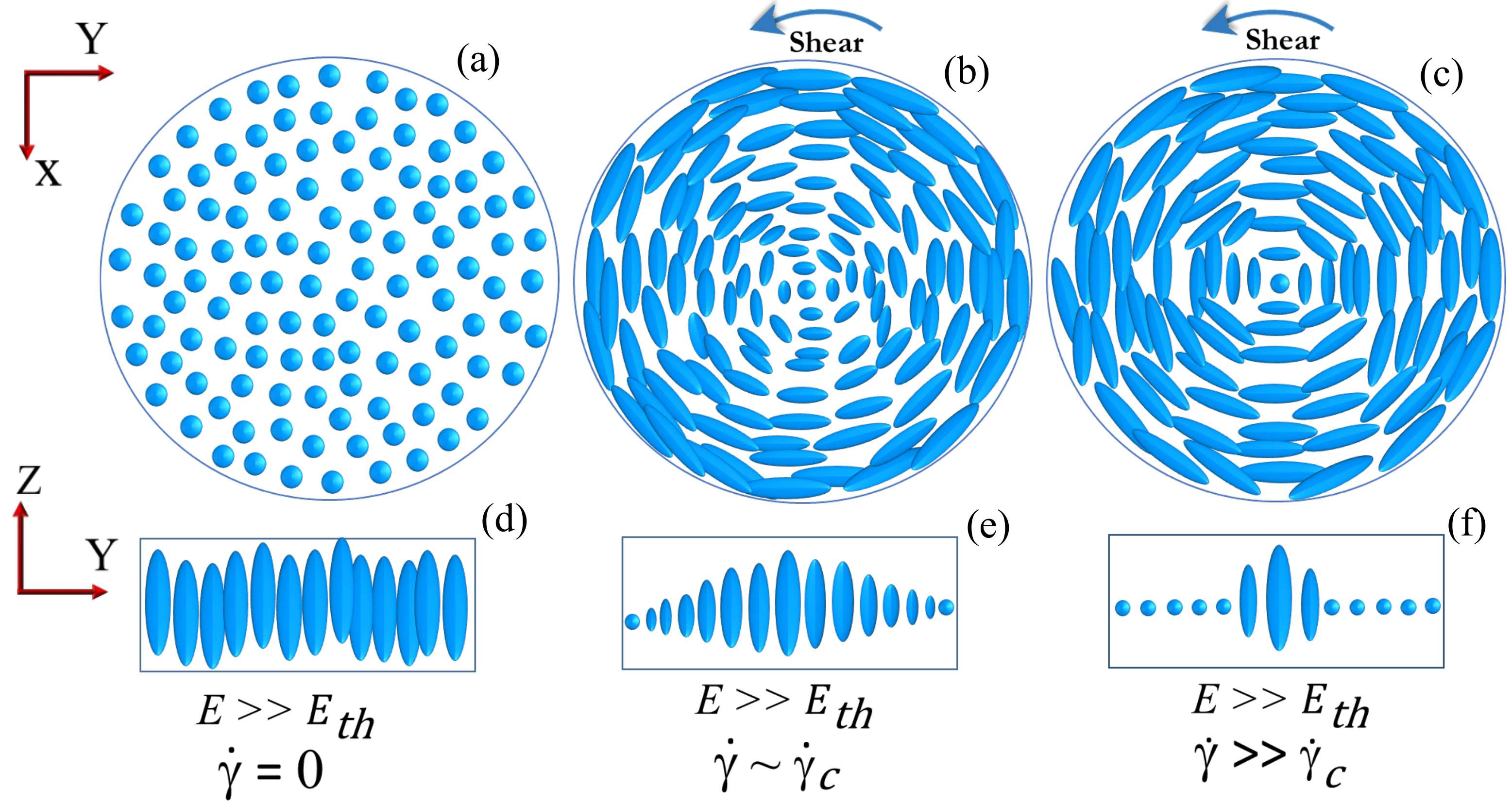}
\caption{Top view (xy-plane) of the director at electric field $E \gg E_{th}$ (along z-axis) and different shear rates, (a) $\dot{\gamma}=0$,  (b) $\dot{\gamma}\sim\dot{\gamma}_{c}$ and (c) $\dot{\gamma} \gg \dot{\gamma}_{c}$. Side views of the director in the vertical mid-plane (zy-plane) at the corresponding shear rates are shown in (d-f). The rotation of the top plate is indicated by blue arrows. Note that the director is twisted radially outward forming a half skyrmion-like structure in (b,e) and the structure is almost uniform except at the centre in figure (c,f).
\label{fig:figure5}}
\end{figure}

 Based on the above results we propose possible director configurations at different shear rates. In our experiment, the bottom plate is fixed and the top plate is rotated with different shear rates. At zero shear rate ($\dot{\gamma}=0$ s\textsuperscript{-1}), the molecules are aligned perpendicular to the plates as the applied electric field is very much greater than the Freedericksz threshold field (see Fig. \ref{fig:figure5}(a)). At a high shear rate ($\dot{\gamma}\gg\dot{\gamma}_c$), the hydrodynamic torque due to the shear flow is much larger than the dielectric torque due to the applied electric field. As a result, the molecules align nearly parallel to the shear flow direction except at the centre as shown in Fig.\ref{fig:figure5}(c) (see later discussion). At the critical shear rate  ($\dot{\gamma}\sim\dot{\gamma}_c$) these two torques are comparable and the director attains an intermediate orientation ($\pi/2\leq\theta\leq\theta_L$).
 
  Further, the shear rate experienced by fluid elements in parallel plate geometry depends on their position $r$ (distance from the centre of plates) and is given by $\dot{\gamma}=r\omega/h$, where $h$ is the gap between the plates and $\omega$ is the angular frequency. Conventionally, in parallel plate systems, the shear rate at the rim ($r=a$) is considered for the calculation of stress and viscosity of Newtonian fluids, where $a$ is the radius of the plate. In complex fluids like liquid crystals, the shear rate, and consequently the hydrodynamic torque very much depends on $r$ and is not the same throughout the plate. The hydrodynamic torque is zero at the centre and it increases from the centre to the edge of the plates. 
   Near the critical shear rate ($0<\gamma\sim\gamma_c$), the director at the centre is vertical and, as one moves towards the perimeter, the hydrodynamic torque increases gradually and the director tilts continuously towards the shear flow direction, creating a half skyrmion-like director deformation as shown in Fig.\ref{fig:figure5}(b). Such a complex director distortion can give rise to flexoelectric polarisation~\cite{blinov,pg} which can couple to the orientational fluctuations of the director. As a result, it can give rise to a collective polarisation mode. However, such polarization is expected to relax at a much lower frequency~\cite{BIO} hence the influence of flexoelectric polarization can be ignored in our experiments.
  The complex director configuration in the low shear rate range ($0<\gamma\leq\gamma_c$) can give rise to a cooperative dipolar motion due to secondary coupling of the steric interactions of the molecules to the electric polarization and consequently reduces the dielectric relaxation frequency. The secondary effect is very different from the primary statistical correlation function associated with the molecular dipole reorientations to polarization, which occurs at a much higher frequency~\cite{BIO}. 
  However, theoretical / simulation studies are required to understand the results quantitatively. 
 
%

 Further, a few comments are in order.  First, two dimensionless numbers, that characterize the flow are Deborah number $D_{e}=\tau\dot{\gamma}$, and Ericksen numbers $E_{r}=\frac{\eta\dot{\gamma}}{K/h^{2}}$~\cite{RGL1},  where $\tau$ is the characteristic relaxation time of the material, $h$ is the gap between the two plates, $\eta$ is the effective viscosity and $K$ is the mean elastic constant. In the experimental shear rate range ($\dot{\gamma}$=1-200 s\textsuperscript{-1}), $E_r=10^3\sim10^5$ and  $D_e=10^{-2}\sim2$ (see Table S2, SM)~\cite{x2}. It suggests liquid-like response and the viscous effects dominate over the elastic effects. 
 Second, liquid crystals contain finite impurity ions which can move towards the opposite electrodes under ac electric field and create space charge polarization. The electrode charging time $\tau=\lambda_D L/2D$, where $\lambda_D$ is the Debye screening length, $L$ is the gap between the electrodes and $D$ is the diffusion constant of the ions. Taking $D\sim10^{-11}$ m\textsuperscript{2}s\textsuperscript{-1}~\cite{sagu}, $\lambda_D\sim0.1~\mu m$~\cite{oleg}, $L=80 ~\mu m$, the estimated electrode polarisation frequency is about 2.5 Hz which is far below the experimental frequency range. Third, since the maximum applied electric field is not very strong (far below the dielectric breakdown) the dielectric response is assumed to be in the linear regime. 

\section{Conclusion} 
The effective dielectric constant and shear viscosity at an electric field higher than the Freedericksz threshold field of a nematic liquid crystal decrease very rapidly under rotational shear due to the tilting of the director in the shear plane from the field to the shear flow direction. The dielectric relaxation frequency shows a pronounced minimum at a critical shear rate. The results suggest the development of a cooperative dipolar relaxation in the low shear rate range. These results are important for applications of liquid crystals as well as polar liquids in electromechanical devices. 
 Considering the competing effects of the hydrodynamic and dielectric torques, we proposed a simple physical model that demonstrates the director configuration at different shear rates qualitatively. In essence, our findings uncover cooperative dipolar relaxation under in the presence of transverse electric and flow fields that reduce dielectric relaxation frequency. We focussed on an apolar nematic LC, however rheo-dielectric studies on liquid crystals with macroscopic polarisations such as ferroelectric nematic and ferroelectric smectics LCs are promising for new relaxation dynamics.

\section{Acknowledgments}
\textbf{Acknowledgments}: SD acknowledges financial support from SERB (SPR/2022/000001). AD acknowledges DST for fellowship (DST/WISE-PhD/PM/2023/57). We thank Dr. Arun Roy for the useful discussions.


\end{document}